\input{aipcheck}

\documentclass[
    ,final            
  ]
  {aipproc}

\layoutstyle{6x9}

\newcommand{\mo}    {M_{\odot}}

\begin{document}

\title{High energy processes in microquasars}

\author{Josep M. Paredes}{
  address={Departament d'Astronomia i Meteorologia, Universitat de
Barcelona, Av. Diagonal 647, 08028 Barcelona, Spain }
}

\begin{abstract}
Microquasars are X-ray binary stars with the capability to generate relativistic
jets. It is expected that microquasars are 
$\gamma$-ray sources, because of the analogy with quasars and because the theoretical models predict emission at such energy range. In addition, from observational arguments, there are two microquasars
that appear as the possible counterparts for two unidentified high-energy $\gamma$-ray sources.
\end{abstract}

\maketitle


\section{Introduction}
Galactic microquasars are certainly one of the most recent
additions to the field of high energy astrophysics and have attracted
increasing interest over the last decade.

The study of microquasars can contribute to an unified understanding of the
accretion and ejection phenomena in the vicinity of collapsed objects
(\cite{mirabel1}), and can help us explain some of the analogous phenomena
observed in distant quasars and active galactic nuclei.
They 
also appear as the possible counterparts for some of the unidentified sources
of high-energy gamma-rays detected by the experiment EGRET on board the
satellite COMPTON-GRO.
However, the $\gamma$-ray 
spectrum of microquasars is the most poorly known, mainly due to the lack of
sensitive instrumentation in the past. Thus, microquasars are now primary targets
for all of the observatories working in the $\gamma$-ray domain.  This paper provides a general review of the main
observational results obtained up to now as well as a summary of the scenarios
for the production of high-energy $\gamma$-rays at the present moment. 

\section{X-ray binaries and microquasars}
An X-ray binary is a binary system containing a compact object, either a
neutron star or a stellar-mass black hole, that emits X-rays as a result of a
process of accretion of matter from the companion star.

In High Mass X-ray Binaries (HMXBs) the donor star is an O or B early type
star of mass in the range $\sim8$--$20$~$M_{\odot}$ and the typical orbital
periods are of several days. HMXBs are conventionally divided into two subgroups:
systems containing a B star with emission lines (Be stars), and systems
containing a supergiant (SG) O or B star. In the first case, the Be stars do
not fill their Roche lobe, and accretion onto the compact object is produced
via mass transfer through a decretion disk. Most of these systems are
transient X-ray sources during periastron passage. In the second case, OB SG
stars, the mass transfer is due to a strong stellar wind and/or to Roche lobe
overflow. The X-ray emission is persistent, and large variability is usual.
The most recent catalogue of HMXBs was compiled by \cite{liu1},
and contains 130 sources.

In Low Mass X-ray Binaries (LMXBs) the donor has a spectral type later than B,
and a mass $\leq2$~$M_{\odot}$. Although it is typically a non-degenerated star,
there are some examples where the donor is a WD. The orbital periods are in
the range 0.2--400 hours, with typical values $<24$ hours. The orbits are
usually circular, and mass transfer is due to Roche lobe overflow. Most of
LMXBs are transients, probably as a result of an instability in the accretion
disk or a mass ejection episode from the companion. The most recent catalogue of LMXBs
was compiled by \cite{liu2} and contains 150 sources.

\begin{figure}[tbp] 
\resizebox{0.8\hsize}{!}{\includegraphics{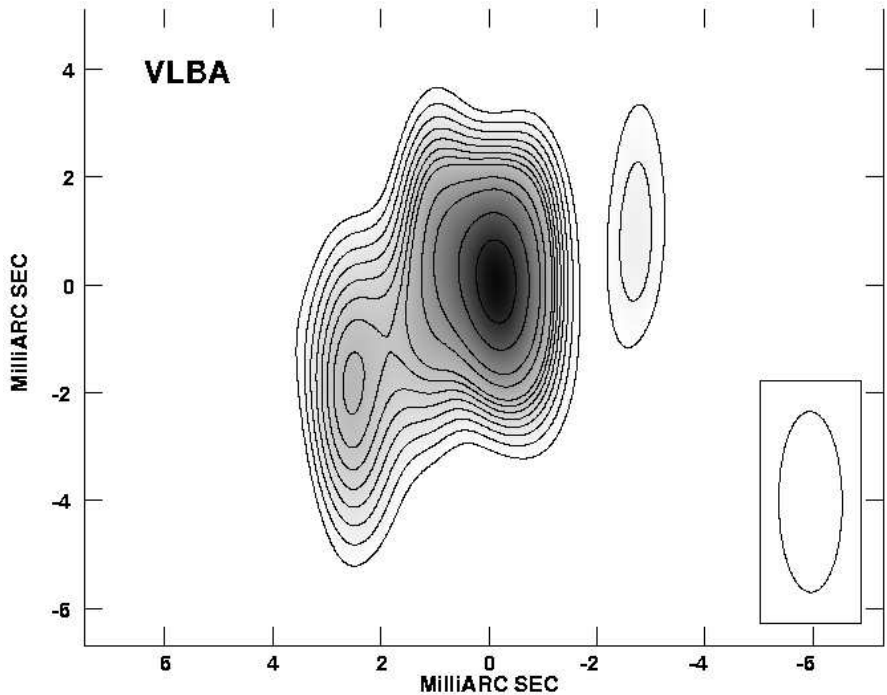}}
  \caption{Relativistic radio jets of LS~5039, observed with the VLBA, which reveal
  its microquasar nature (\cite{paredes1}).}
  \label{vlbajet}
\end{figure}

The first X-ray binary known to display radio emission was Sco~X-1 in the late
1960s. Since then, many X-ray binaries have been detected at radio wavelengths (REXBs). Considering both
catalogues together, there are a total of 43 radio emitting sources (8 HMXBs and 35
LMXBs)(\cite{ribo1}). Some of these REXBs, those that show a relativistic radio jet,
are named microquasars (\cite{mirabel1}). In Fig.~\ref{vlbajet}, we show the 
relativistic jets of the microquasar LS~5039.

The concept of microquasar has been widely accepted in recent years as a new kind of
X-ray binary stars in our Galaxy with the capability to generate collimated beams, or
jets, of relativistic plasma. The ejection takes place in a bipolar way from the
accretion disk associated with the compact star, a black hole or a neutron star. The
word microquasar itself was chosen by the analogy of these astronomical objects with
quasars and other active galactic nuclei (AGNs) at cosmological distances
(\cite{mirabel1}).
The analogy quasar-microquasar goes beyond a simple morphological resemblance. 
Today, there is growing evidence to think that the physics involved in both types of 
objects is the same, or at least very similar. The key difference would be the 
distinct order of magnitude of the most significant parameters, especially the 
mass of the compact object.

The current number of microquasars is $\sim$15 among the 43 catalogued REXBs 
(\cite{ribo1}). Some authors (\cite{fender1}; \cite{fender2}) have proposed 
that all REXBs are microquasars and
would be detected as such provided that there is enough sensitivity and/or
resolution in the radio observations. The known microquasars, compiled from
different sources, are listed in Table~\ref{census}, where  
the following information is given: name; type of
system; distance; orbital period; mass of the compact object; degree of
activity (persistent/transient radio emission); and apparent velocity of the
ejecta. Extensive reviews on microquasars can be found in \cite{mirabel1}, 
\cite{fender2} and \cite{ribo2}.

\begin{table}[!t]
\begin{tabular}{lcccccc}
\hline
  \tablehead{1}{l}{b}{Name \\} &
  \tablehead{1}{c}{b}{System\\type\tablenote{NS: neutron star; BH: black hole}} &
  \tablehead{1}{c}{b}{$D$\\(kpc)} &
  \tablehead{1}{c}{b}{$P_{\rm orb}$\\(d)} &
  \tablehead{1}{c}{b}{$M_{\rm compact}$\\ $\mo$} & 
  \tablehead{1}{c}{b}{Activity\\radio\tablenote{p:persistent; t: transient}} &
  \tablehead{1}{c}{b}{$\beta_{\rm apar}$ \\} \\
\hline
\tablehead{7}{c}{c}{\bf High Mass X-ray Binaries (HMXB)} \\
\hline
{\bf LS~I~+61~303}  &B0V & 2.0 & 26.5 & $-$ & p & $\geq$0.4  \\
& +NS? &  & & & & \\

{\bf V4641~Sgr}   & B9III & $\sim10$ & 2.8  & 9.6 & t & $\ge9.5$ \\
 &+BH \\
 
{\bf LS~5039}  &  O6.5V((f)) & 2.9 & 4.4 & 1$-$3 & p  & $\geq0.15$ \\
 & +NS?& &   & &   \\
  
{\bf SS~433} &evolved A?   & 4.8  &  13.1 &  11$\pm$5?& p & 0.26      \\
 &+BH?  & &   & & & \\

{\bf Cygnus~X-1} &O9.7Iab  & 2.5 & 5.6 & 10.1 &  p  \\
& +BH & &   & &  \\
 
{\bf Cygnus~X-3} &WNe     &  9      &  0.2   & $-$& p  & 0.69     \\
& +BH? & &   & &  \\

\hline
\tablehead{7}{c}{c}{\bf Low Mass X-ray Binaries (LMXB)} \\
\hline
      
{\bf Circinus~X-1}    & Subgiant &  5.5     &  16.6  &$-$&  t   & $>15$ \\
 & +NS \\
 
{\bf XTE~J1550$-$564} & G8$-$K5V & 5.3  & 1.5   & 9.4 &  t
&$>2$  \\
& +BH & &   & &  \\
 
{\bf Scorpius~X-1}   & Subgiant    & 2.8      &  0.8  &  1.4 &  p    \\
 & +NS  \\
  
{\bf GRO~J1655$-$40}  & F5IV  & 3.2    &  2.6   & 7.02 & t & 1.1    \\
& +BH & &   & &    \\
   
{\bf GX~339$-$4}   & $-$         & $>6$     &  1.76  & 5.8$\pm$0.5 & t& $-$  \\
& +BH \\

{\bf 1E~1740.7$-$2942}&$-$  & 8.5? &  12.5?   &$-$ & p &$-$ \\
& +BH ?& &   & & &  \\

{\bf XTE~J1748$-$288} & $-$ &  $\geq8$   &  ?    &   $>4.5$? & t & 1.3        \\
 &+BH? \\
    
{\bf GRS~1758$-$258}  & $-$ & 8.5?  & 18.5?  &$-$   & p &$-$\\
&+BH ?\\
   
{\bf GRS~1915+105}&  K$-$M III   & 12.5 & 33.5  & 14$\pm$4 &  t & 1.2$-$1.7 \\
 &+BH & &   & &  \\
\hline
\end{tabular}
\caption{Microquasars in our Galaxy}
\label{census}
\end{table}

\section{Models for $\gamma$-ray production in microquasars}

It is widely accepted that relativistic jets in AGNs are strong emitters of
$\gamma$-rays with GeV energies (\cite{montigny1}).
Generally speaking, and allowing for their similarity (\cite{mirabel1}), 
one could also expect the jets in microquasars
to be GeV $\gamma$-ray emitters. In some cases, however, the sensitivity of
the current $\gamma$-ray detectors may not be high enough to detect such
emission. For instance, based on the physical parameters derived from
observations of outbursts, the expected $\gamma$-ray flux of GRS~1915+105 up
to very high-energy $\gamma$-rays has been estimated from inverse Compton
scattering of the synchrotron photons (\cite{atoyan1}). The
resulting fluxes could have been hardly detected by EGRET, being also of
transient nature, but they are within the sensitivity of the future AGILE and
GLAST missions, about 10--100 times better than that of EGRET. 

Several models have been developed to explore the high energy emission from
the jets of microquasars. Two kinds of models can be found in the literature
depending on whether hadronic or leptonic jet matter dominates the emission at
such an energy range: the hadronic jet models (\cite{romero1}; \cite{bosch1}), 
and the leptonic jet models. Among leptonic jet models, there
are IC jet emission models that can produce X-rays and $\gamma$-rays, based in
some cases on the synchrotron self-Compton (SSC) process (i.e. \cite{grindlay1}; 
\cite{atoyan1}), and in
other cases on external sources for the IC seed photons (EC) 
(i.e. \cite{kaufman1}; \cite{georganopoulos1}). 
In addition, there are synchrotron jet emission
models that can produce X-rays (i.e. \cite{markoff1}). A general
description of such models can be found in \cite{romero2}.

\section{Microquasars as low energy $\gamma$-ray sources}

\cite{bird1} have reported the first high-energy survey catalog
obtained with the IBIS $\gamma$-ray imager on board INTEGRAL, covering the
first year data. This initial survey has revealed the presence of $\sim$120
sources detected with a good sensitivity in the energy range 20$-$100~keV.
Among the detected sources, we have inspected the microquasars listed in
Table~\ref{census}. In the second column of Table~\ref{detections}, we list
their flux (count/s) and error or upper limit in the energy range of
40$-$100~keV.

The Burst and Transient Source Experiment (BATSE), aboard the Compton Gamma
Ray Observatory (CGRO), monitored the high energy sky using the Earth
occultation technique (EOT). A compilation of BATSE EOT observations has been
published recently (\cite{harmon1}), with the flux data for the
sample being presented in four energy bands. From this catalog we have
also selected the data on microquasars. In the third column of
Table~\ref{detections}, we have listed their flux in the energy
range 160$-$430~keV in mCrab units. Cygnus~X-1 and Cygnus~X-3 have been
studied extensively by BATSE.

The instrument COMPTEL, also aboard the CGRO, detected 32 steady sources and
31 $\gamma$-ray bursters (\cite{schonfelder1}). Among the
continuum sources detected there are the microquasar Cygnus~X-1 and other two
sources, GRO~J1823$-$12 and GRO~J0241+6119, possibly associated with two other
microquasars (see the fourth column in Table~\ref{detections}). 

The standard interpretation of the emission in the low-energy $\gamma$-ray
range is that disc blackbody photons are Comptonized by thermal/nonthermal
electrons. There are state transitions (hard and low states) thought to be
related to changes in the mass accretion rate. Nevertheless, it is still
unclear whether this is what really happens. Alternatively, some groups have
suggested that this emission could come from the jet, based on
recent observational and theoretical results 
(see, i.e., \cite{fender3}; \cite{markoff1}; \cite{georganopoulos1}).

\begin{table}[!t]
\begin{tabular}{lcccc}
\hline
 \tablehead{1}{l}{b}{Name\\  \\} &
 \tablehead{1}{c}{b}{INTEGRAL\tablenote{The first IBIS/ISGRI 
 soft gamma-ray galactic plane survey catalog 
(\cite{bird1}).}\\40$-$100 keV\\(count/s)} &
 \tablehead{1}{c}{b}{BATSE\tablenote{BATSE Earth occultation catalog, Deep 
 sample results 
(\cite{harmon1}).}\\160$-$430 keV \\   (mCrab)}  &
 \tablehead{1}{c}{b}{COMPTEL\tablenote{The first COMPTEL source catalogue 
 (\cite{schonfelder1})}\\
  1$-$30 MeV \\ (GRO) } &
 \tablehead{1}{c}{b}{EGRET\tablenote{The third EGRET catalog of 
 high-energy $\gamma$-ray sources (\cite{hartman1})}\\ $>$ 100 MeV \\ (3EG)} \\
\hline
\tablehead{5}{c}{b}{\bf High Mass X-ray Binaries (HMXB)} \\
\hline

{\bf LS~I~+61~303} &  $-$  & 5.1$\pm$2.1  & J0241+6119? & J0241+6103?\\

{\bf V4641~Sgr}  & $-$ & $-$   & $-$ & $-$ \\

{\bf LS~5039} &  $-$ &  3.7$\pm$1.8 & J1823$-$12?  & J1824$-$1514? \\
  
{\bf SS~433} & $<$1.02  &  0.0$\pm$2.8 & $-$ & $-$  \\
  
{\bf Cygnus~X-1}  & 66.4$\pm$0.1  & 924.5$\pm$2.5 &  yes &$-$\\
  
{\bf Cygnus~X-3}    &  5.7$\pm$0.1     & 15.5$\pm$2.1 & $-$   & $-$ \\

\hline
\tablehead{5}{c}{b}{\bf Low Mass X-ray Binaries (LMXB)}\\
\hline
      
{\bf Circinus~X-1}    &  $-$     &   0.3$\pm$2.6&  $-$   & $-$\\
 
{\bf XTE~J1550$-$564} &  0.6$\pm$0.07    & $-$2.3$\pm$2.5 & $-$   & $-$ \\
 
{\bf Scorpius~X-1}     &  2.3$\pm$0.1      &  9.9$\pm$2.2 &  $-$ & $-$  \\
  
{\bf GRO~J1655$-$40} & $-$    &   23.4$\pm$3.9 & $-$ & $-$  \\
   
{\bf GX~339$-$4}   &   0.55$\pm$0.03    &   580$\pm$3.5 & $-$&$-$  \\ 
  
{\bf 1E~1740.7$-$2942}&  4.32$\pm$0.03 &  61.2$\pm$3.7 & $-$& $-$  \\
 
{\bf XTE~J1748$-$288} &   $-$  &    $-$ & $-$ & $-$        \\

{\bf GRS~1758$-$258}  &  3.92$\pm$0.03  &  38.0$\pm$3.0  & $-$ &$-$\\
   
{\bf GRS~1915+105}    &   8.63$\pm$0.13 &  33.5$\pm$2.7 & $-$  & $-$ \\
\hline
\end{tabular}
\caption{High energy emission from microquasars}
\label{detections}
\end{table}

\section{Microquasars as high energy $\gamma$-ray sources}

\begin{figure}[tbp] 
\resizebox{0.8\hsize}{!}{\includegraphics{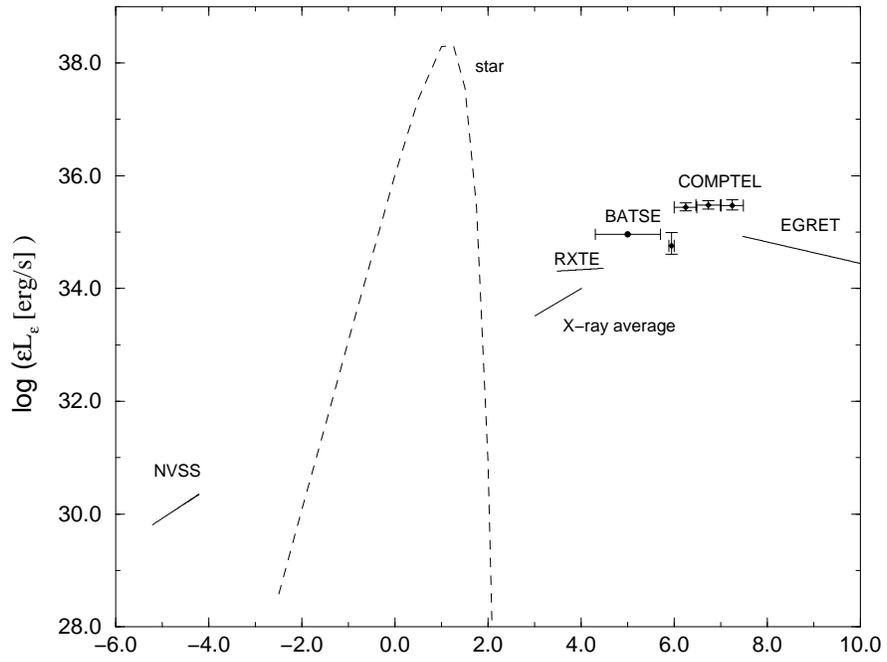}}
  \caption{Observed spectral energy distribution of LS~5039.}
  \label{ls5039}
\end{figure}

\begin{figure}[tbp] 
\resizebox{0.8\hsize}{!}{\includegraphics{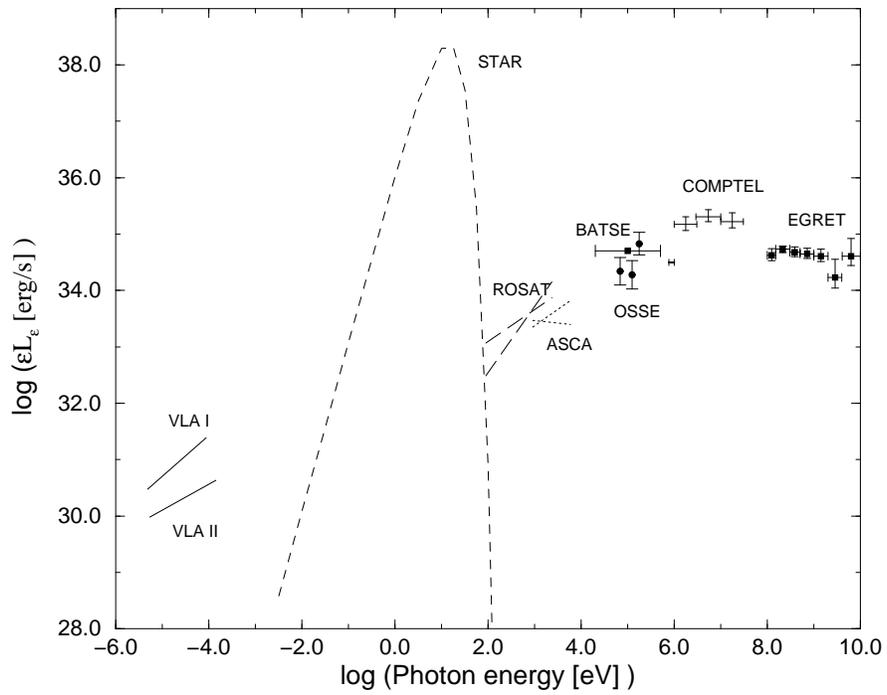}} 
  \caption{Observed spectral energy distribution of LS~I~+61~303.}
  \label{lsi61303}
\end{figure}

Up to now, there are  
two HMXB microquasars, LS~5039 and LS~I~+61~303, that are associated with
two EGRET sources. Both sources have been extensively studied at different 
wavelengths, and both have a very similar spectral energy distribution as can
be seen in Figure~\ref{ls5039} and Figure~\ref{lsi61303}.

\subsubsection{LS~5039 / 3EG~J1824$-$1514}

The discovery of the microquasar LS~5039, and its possible association with a
high-energy $\gamma$-ray source ($E>$100~MeV), provides observational evidence
that microquasars could also be sources of high-energy $\gamma$-rays 
(\cite{paredes1}). It is important to point out that this was the first time
that an association between a microquasar and a high-energy $\gamma$-ray
source was reported. This finding opened up the possibility that other
unidentified EGRET sources could also be microquasars. LS~5039 is the only
X-ray source from the bright ROSAT catalogue whose position is consistent with
the high energy $\gamma$-ray source 3EG~J1824$-$1514. LS~5039 is also the only
object simultaneously detected in X-rays and radio 
which displays bipolar radio jets at sub-arcsecond scales.
New observations conducted with the EVN and MERLIN
confirm the presence of an asymmetric two-sided jet reaching up to
$\sim$1000~AU on the longest jet arm 
(\cite{paredes2}, \cite{ribo1}). 

Recently, \cite{collmar1} has reported the detection of an unidentified
$\gamma$-ray source, GRO~J1823$-$12, at galactic coordinates
($l$=17.5$^{\circ}$, b=$-$0.5$^{\circ}$) by the COMPTEL experiment. 
This source is among the strongest COMPTEL sources.
The source region, detected at a high significance level, contains several
possible counterparts, one of which is LS~5039. It is also worth noting that
BATSE has detected this source at soft $\gamma$-rays (see
Table~\ref{detections}). Taking into account these observational evidences,
from radio to high-energy $\gamma$-rays, LS~5039 appears to be a very likely
counterpart of the EGRET source 3EG~J1824$-$1514.

The $\gamma$-ray emission from 3EG~J1824$-$1514, with a luminosity of
L$_{\gamma} (>$100~MeV) $\sim$ 10$^{35}$ erg~s$^{-1}$, is likely to originate
from inverse Compton effect of the ultraviolet photons from a hot companion
star scattered by the same relativistic electrons responsible for the radio
emission. The energy shift in this process is given by
E$_\gamma\sim\gamma_{\rm e}^{2}$ E$_{ph}$, where the energies of the
$\gamma$-ray and the stellar photon are related by the Lorentz factor of the
electrons squared. For an O6.5 star in the main sequence, such as the
component of LS~5039, most of its luminosity is radiated by photons with
E$_{ph}\sim$10~eV. In order to scatter them into $\gamma$-ray photons with
E$_{\gamma}\sim$100~MeV, electrons with a Lorentz factor of $\gamma_{\rm
e}\sim$10$^4$, or equivalently with energy $\sim$10$^{-2}$~erg, are required.

Recently, \cite{bosch2} have explored with a detailed numerical model if this
system can both produce the emission and present the variability detected by
EGRET ($>$100~MeV), and obtained positive results. 

\subsubsection{LS~I~+61~303 / 3EG~J0241+6103}

After the discovery of relativistic jets in LS~I~+61~303, this source 
has been classified as a new microquasar (\cite{massi1}, 
\cite{massi2}). This object has also been
proposed to be associated with the $\gamma$-ray source 2CG~135+01
(=3EG~J0241+6103) (\cite{gregory1}; \cite{kniffen1}). 
Although the broadband 1~keV--100~MeV spectrum of
LS~I~+61~303 remains uncertain, because OSSE and COMPTEL observations were
likely dominated by the quasar QSO~0241+622 emission, the EGRET angular
resolution is high enough to exclude this quasar as the source of the
high-energy $\gamma$-ray emission (\cite{harrison1}).  BATSE marginally
detected the source, and the quasar was also excluded as the origin of this
emission (see Table~\ref{detections}). The proposed association between LS~I~+61~303 and the high-energy
$\gamma$-ray source is still unclear due to the low angular resolution of EGRET,
although no radio loud active galactic nucleus or strong radio pulsar is known
within the $\gamma$-ray error box, which includes LS~I~+61~303 (\cite{kniffen1}).

Recently, \cite{massi3} has carried out a timing analysis of pointed
EGRET observations (\cite{tavani1})
suggesting a period of 27.4$\pm$7.2 days, in agreement with the orbital period
of this binary system of 26.496 days. This result, if confirmed, would
clearly support the association of LS~I~+61~303 with 3EG~J0241+6103.

This microquasar also seems to be a fast precessing system. MERLIN images
obtained in two consecutive days show a change in the
direction of the jets of about 50$^{\circ}$ that has been interpreted as a
fast precession of the system (\cite{massi2}). If this is
confirmed, it could solve the puzzling VLBI structures observed so far, as
well as the short term variability of the associated $\gamma$-ray source
3EG~J0241+6103. 

Up to now, the only existing radial velocity curve of LS~I~+61~303 was that
obtained by \cite{hutchings1}. Recently, after a
spectroscopic campaign, an improved estimation of the orbital parameters has
been obtained (\cite{casares1}). Here, we will just mention the
new high eccentricity (e=0.72$\pm$0.15) and the periastron orbital phase at
$\sim$0.2. These values are a key information for any interpretation of the
data obtained at any wavelength.

Several models have 
been proposed to explore the high
energy emission of this source (e.g. \cite{taylor1};
\cite{punsly1}; \cite{harrison1}; 
\cite{leahy1}). The
most recent theoretical work has been presented by \cite{bosch3}, 
who explore with a  detailed numerical model if this
system can both produce the emission and present the variability detected by
EGRET ($>$100~MeV), and obtained positive results.  

\begin{figure}[tbp] 
\resizebox{0.8\hsize}{!}{\includegraphics{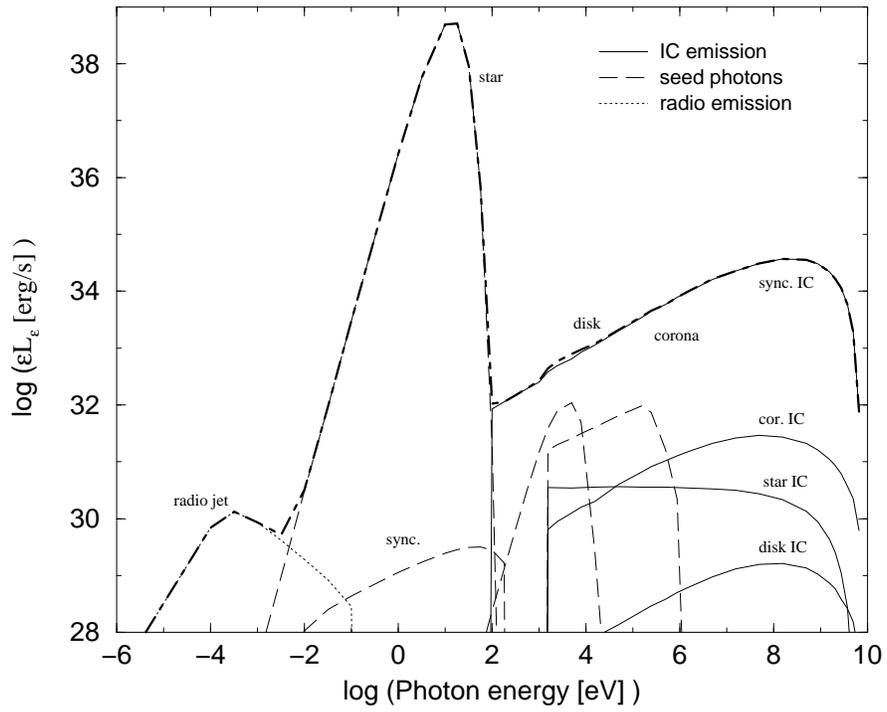}}
  \caption{Spectral energy distribution model of a high mass microquasar 
  (\cite{bosch2}).}
  \label{model}
\end{figure}

\begin{figure}[tbp]
\resizebox{0.8\hsize}{!}{\includegraphics{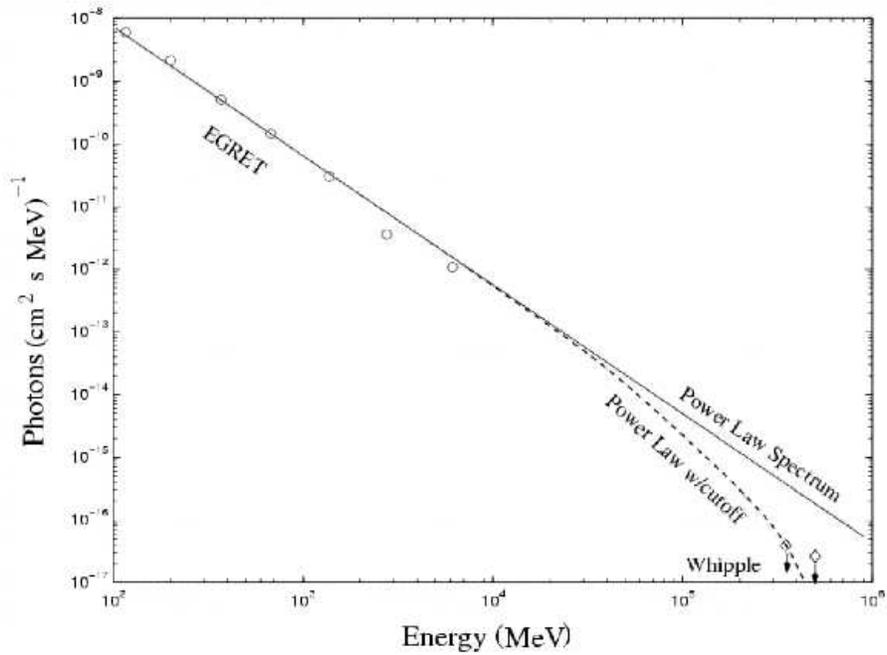}}
  \caption{{\small EGRET data points of 3EG~J0241+6103 (circles) and the flux 
  upper limits
   (diamonds) obtained by Whipple (\cite{hall1}).}}
  \label{whipple}
\end{figure}

\subsubsection{High mass microquasars}

Population studies of unidentified EGRET sources suggest that there exists
a group formed by young objects distributed along the galactic plane that
display clear $\gamma$-ray emission variability on timescales from days to months. 
\cite{bosch2} have suggested that these sources 
might be HMXB microquasars and have elaborated some detailed models,
that include both external and SSC scattering, to explain the broad-band
spectrum all the way up to high-energy $\gamma$-rays. It gives further support
to the proposed association between microquasars and  $\gamma$-ray sources.

The computed spectral energy distribution of an EGRET source high-mass 
microquasar is presented in Figure~\ref{model}. Looking at the 
Figures~\ref{ls5039}~and~\ref{lsi61303}, and comparing them with 
Figure~\ref{model}, it is seen how the IC jet scenario reproduces pretty 
well the data, giving further suport to the proposal of microquasars as 
$\gamma$-ray sources.

\section{Microquasars as very high-energy $\gamma$-ray sources}

Regarding the physical scenarios where very high-energy $\gamma$-rays could be
produced, I expose briefly the general characteristics of hadronic and
leptonic jet models. In the context of hadronic jet models, I focus on the
work of \cite{romero1}, where the pion-decay emission produced by
interaction between very high energy protons of a jet (100~TeV) and
environment {\it cold} protons was studied. What is relevant here is the lack
of steepening in the energy spectrum of protons, which suffer much less
energy losses than electrons, implying a quite hard spectrum of the emission
produced by such a mechanism. Therefore, if there is an important population
of very high energy protons, significant emission at very high-energy
$\gamma$-rays presenting a {\it hard} spectrum would be expected. Otherwise,
in the context of leptonic jet models, if electrons with energies of the order
of TeV are present in the jet, emission up to few hundreds of GeV is predicted
(see \cite{atoyan1}; \cite{bosch4}). 
In this case, however, the GeV--TeV spectrum should be quite
softer than the one at lower energies, and also softer than the one expected
in the hadronic case. The determination of the spectrum of a microquasar at
GeV-TeV energies (i.e. photon index, maximum energy, turning point in the
slope) would give us information in order to constrain the models of
acceleration of particles, jet matter nature, and jet physical conditions in
general. These issues will not only improve our knowledge at very high
energies, but also in the entire spectrum, going deeper in our knowledge about
microquasars. 

The observed very high energy (VHE) sky map contains a reduced number of sources. The number
of confirmed and probable catalogued sources at present amounts to fourteen (6 AGN, 3
pulsar wind nebulae, 3 supernova remnants, 1 starburst galaxy, and 1 unknown)
(\cite{ong1}). Some microquasars have been observed in the energy range
of TeV $\gamma$-rays with the imaging atmospheric Cherenkov telescopes, but
none of them has been detected with high confidence yet. Historically,
Cygnus~X-3 was widely observed with the first generation of TeV instruments. 
Some groups claimed that they had detected Cygnus~X-3 (\cite{chadwick1}) 
whereas other groups failed to detect it (\cite{flaherty1}). 
As the claimed detections have not been confirmed, and
the instrumentation at this epoch was limited, these results have not been
considered as positive detections by the astronomical community. The HEGRA
experiment detected a flux of the order of 0.25~Crab from GRS~1915+105 during
the period May-July 1996 when the source was in an active state 
(\cite{aharonian1}). This source has also been observed with Whipple,
obtaining a 3.1$\sigma$ significance (\cite{rovero1}). More
recently, an upper-limit of 0.35 Crab above 400~GeV has been quoted for 
GRS~1915+105 (\cite{horan1}). However, these results
need further confirmation given the marginal confidence of the detection.
LS~I~+61~303 was observed too, but was not detected in the TeV energy range 
(\cite{hall1}). The results, obtained by Whipple at two epochs, show flux upper
limits for LS~I~+61~303 (see Figure~\ref{whipple}) implying a steepening of 
the spectrum beyond EGRET
energies. The fact that there seems to be a steepening seems to make
the leptonic model more likely than the hadronic one.

\section{Summary}

The standard physical scenario for microquasars predicts them to be sources
of low and high energy
$\gamma$-ray emission. Actually, most microquasars have been detected at
soft $\gamma$-ray by several instruments on board satellites. At higher energies,
microquasars appear as a possible explanation for some of the unidentified
high energy $\gamma$-ray sources detected by the experiment EGRET on board
the satellite COMPTON-GRO. The possible association between the microquasar
LS~5039 and the high-energy $\gamma$-ray source 3EG~J1824$-$1514
provides observational evidence that microquasars could also be sources of
high energy $\gamma$-rays (\cite{paredes1}). LS~I~+61~303 has
also been proposed to be associated with the $\gamma$-ray source 2CG~135+01
(=3EG~J0241+6103) (\cite{kniffen1}). Future missions (AGILE, GLAST)
will confirm or reject the proposed association between some microquasars and
EGRET sources. 

From the
theoretical point of view, microquasars might emit at hundreds of GeV and
beyond. Nevertheless, nearly half of the reduced number of known VHE sources 
belong to our Galaxy but none of them is a microquasar. The fact that up to now 
there have not been
reliable detections of microquasars, could be more related to the technical
development of previous instruments than to the physical capability of
microquasars to emit at hundreds of GeV. 
Observations are not yet conclusive in any way as to whether microquasars
emit at such energies. The new generation of Cherenkov imaging telescopes
can help clarify this point.


\begin{theacknowledgments}
I acknowledge partial support by DGI of the Ministerio de Ciencia y
Tecnolog\'{\i}a (Spain) under grant AYA2001-3092, as well as partial support
by the European Regional Development Fund (ERDF/FEDER).  
\end{theacknowledgments}





{}


\end{document}